\documentclass[final,5p,times,twocolumn]{elsarticle}

\usepackage{amssymb}
\usepackage{hyperref}

\journal{Astronomy and Computing}

\begin{document}

\begin{frontmatter}



\title{Mitigating Satellite Trails: a Study of Residual Light after Masking}


\author[inst1]{Imran Hasan \fnref{label1}}
\fntext[label1]{Present address: Nielsen Corp., 85 Broad St., New York, NY 10004}

\affiliation[inst1]{organization={Department of Physics and Astronomy, University of California},
            addressline={One Shields Ave.}, 
            city={Davis},
            postcode={95616}, 
            state={CA},
            country={USA}}

\author[inst1]{J. Anthony Tyson\corref{cor1}}
\cortext[cor1]{Corresponding author}
\ead{tyson@physics.ucdavis.edu}

\author[inst2]{Clare Saunders}

\affiliation[inst2]{organization={Department of Astrophysical Sciences},
            addressline={4 Ivy Lane}, 
            city={Princeton},
            postcode={08544}, 
            state={NJ},
            country={USA}}

\author[inst3]{Bo Xin \fnref{label2}}
\fntext[label2]{Present address: Giant Magellan Telescope Organization, 465 N Halstead St., Pasadena, CA 91107}

\affiliation[inst3]{organization={Rubin Observatory},
            addressline={950 N. Cherry Ave.}, 
            city={Tucson},
            postcode={85719}, 
            state={AZ},
            country={USA}}

\begin{abstract}
Using Hyper Suprime Camera data (a precursor of what is to come with Rubin Observatory) we assess trail masking mitigation strategies for satellite contamination. We examine HSC data of the Hubble COSMOS field where satellite trails have been identified by eye. Exercising the current LSST Science Pipelines on this data, we study the efficacy of masking satellite trails which appear in single visit exposures before they are assembled into a coadded frame. We find that the current routines largely mask satellite trails in single visits, but miss the extended low surface brightness features of the satellite trails. For a sufficiently wide mask, these faint features appear at a less significant level in the final coadd, as they are averaged down in a stack of tens of exposures.  We study this print-through vs mask width. In this note, we describe some of the challenges we encountered in that effort, prospects for more complete removal of the low surface brightness tails of the masked trails, and possible science impacts.
\end{abstract}




\end{frontmatter}



\section{Introduction}
Some 64 years ago, Sputnik opened a new theater for humanity's reach. Today, thousands of commercial low earth orbit satellites (LEOsats) orbit the earth, providing communication and imaging services. This number is expected to increase by two orders of magnitude in the coming decade~\citep{mcdowell-2020, walker+hall2020}. This increase will be driven primarily by large constellations of satellites, operated world wide. The proliferation of satellites will alter the contents of the night sky, impacting astronomy from star gazing to major optical astronomy facilities. The impact of LEOsats on Rubin Observatory's Legacy Survey of Space and Time (LSST) data and possible mitigation was explored in \citep{tyson2020mitigation}. There are two regimes of image contamination from bright satellite trails: (1) for satellites brighter than about 7th $g$ magnitude, the induced non-linear crosstalk on the CCD detectors is difficult to remove, and (2) even after mitigating crosstalk, the main trail itself presents challenges due to residual false alerts and low level systematics. One example of the effect of incompletely suppressed trails at ultra low surface brightness is a systematic weak lens cosmic shear signal from lines of correlated noise. This motivates the current study. 


\section{Methods}
To examine the impact LEOsats will have on Rubin Observatory LSST images and possible mitigation strategies, we turn to a Rubin Observatory precursor survey, the Subaru Hyper Suprime Camera (HSC). Specifically, we utilize observations that overlap the Hubble COSMOS field. We use a catalogue of visits (a telescope pointing to a sky region) and patch numbers for which satellite trails were identified \citep{Saunders2020}.
The catalogue is accurate, though not necessarily complete, and offers an ideal setting to study satellite trails in real observational data.

\subsection{Examining trails on Single Visit Warps}

In the LSST Science Pipelines, individual exposures from chips are re-sampled onto a common coordinate system to create $warps$. The warps are then used as input to create a $coadds$ whose pixels are assembled via the process described below. Currently, the \texttt{CompareWarpAssembleCoaddTask} in the science pipelines v2.1 is being used to assemble coadded frames from warps. Heuristically, this algorithm performs difference imaging between single exposure warps and aggressively clipped coadd of many exposures to identify transient features. Detected transient features are clipped, and their corresponding pixels during coaddition are flagged with the CLIPPED bit mask so they do not propagate to the final coadded image. 

There is also a routine that identifies morphological features that resemble straight lines in the data by performing a kernel-based Hough transform on the images \citep{fernandes2008}, and likewise masks their pixels during coaddition so that the masked area does not contribute to the final coadded image.

\begin{figure}
	\includegraphics[width=\columnwidth]{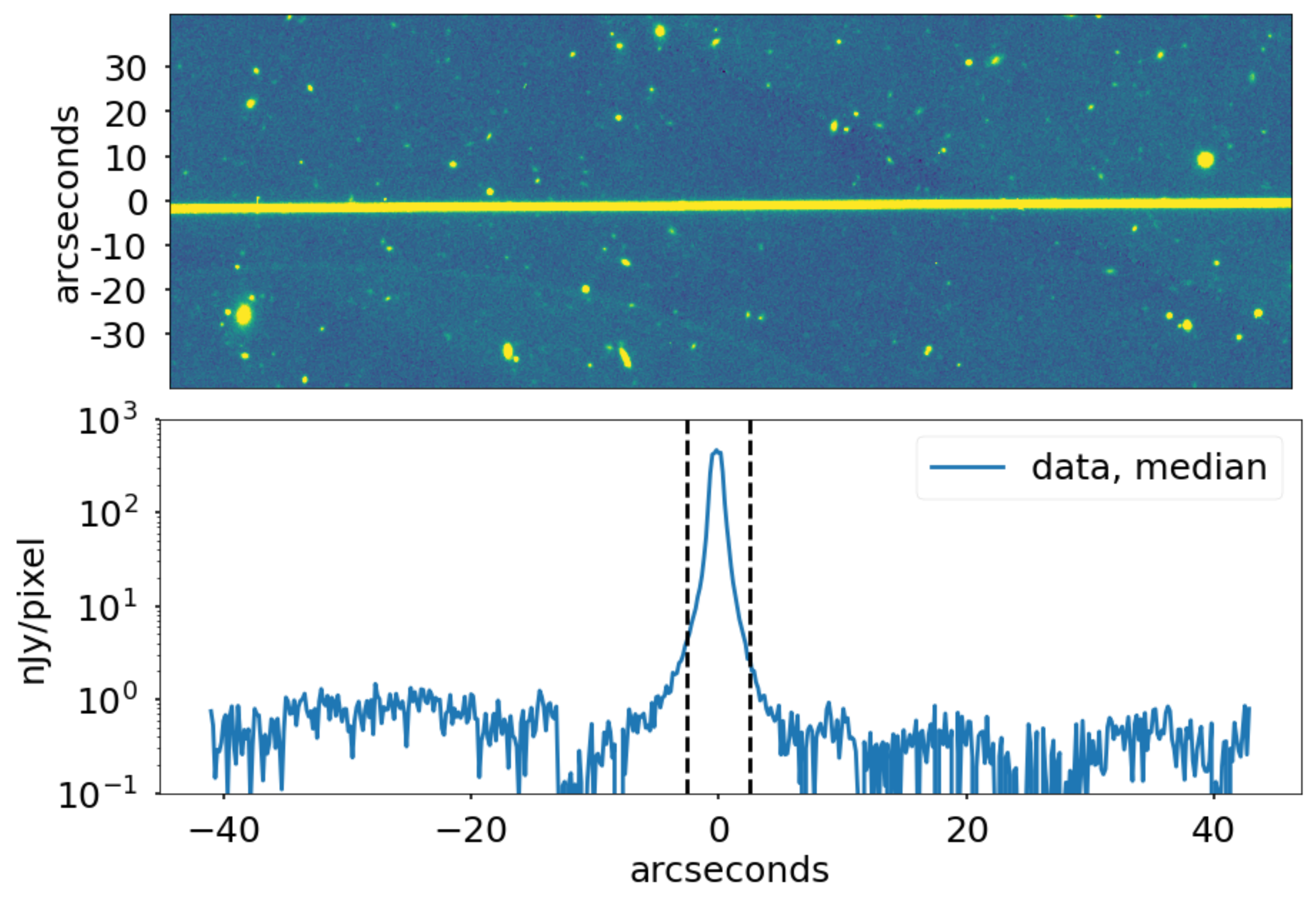}
    \caption{\emph{Top}: A section of a re-sampled single exposure warp from HSC observations of the COSMOS field where a satellite contaminated the exposure. Also pictured are some background galaxies throughout the image, and an optical ghost in the bottom left of the image. \emph{Bottom}: median surface brightness profile of the re-sampled warp in units of nJy/pixel, where we have collapsed the signal in the top panel along the x axis. The surface brightness profiles peak at the location of the satellite trail in the image. In dashed black lines we show the approximate extent of the mask generated by the coaddition process. The mask largely clips out the satellite trail and part of its extended lower surface brightness tails, although some of the low surface brightness tails 'bleed' out from beneath the mask. }
    \label{fig:warp}
\end{figure}

We examine a warp after it has been processed through to coaddition, and has had clipping and masking performed on it as a result. In this case, the CLIPPED mask plane bit is set over the satellite trail. This happens if the feature is clipped before the trail masking step, which occurs often due to the temporal filtering employed by the LSST Science Pipelines. In the top panel of Figure \ref{fig:warp} we show a zoom in on a single visit re-sampled warp which has been contaminated by a satellite trail. In the bottom panel of the Figure, we show the surface brightness profile of the satellite trail in the blue curve, where we have collapsed the signal in the top panel along the x-axis. The bottom panel also includes vertical dashed black lines which approximately represent the width of the mask that is ultimately generated for the satellite trail. The mask covers most of the trail's profile except for the outermost wings before they blend into the astrophysical background. As discussed in \citep{tyson2020mitigation}, a typical 27 g magnitude faint galaxy in the LSST 'Gold Sample' is $\sim$57 nJy integrated flux. This flux is spread over roughly 1 square arcsecond by the convolution of the PSF with the galaxy.  At the 0.2 arcsecond pixel scale of LSST this 27th mag galaxy has a surface brightness of 2 nJ/pixel.

\begin{figure}
    \centering
    \includegraphics[width=\columnwidth]{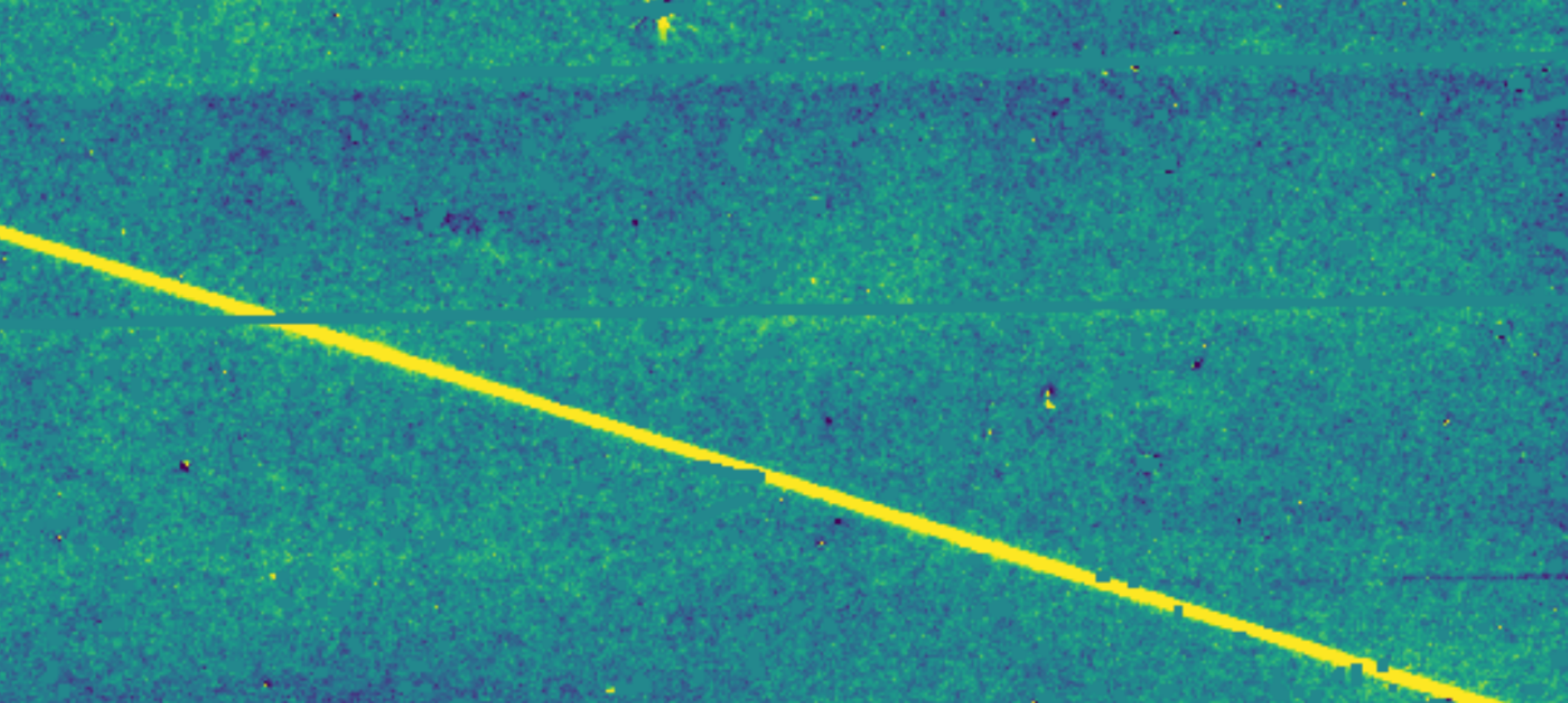}
    \caption{A difference image between a single visit warp and a template coadd, made from aggressively clipping and coadding many frames. The satellite trail is readily recognizable by eye, and detected algorithmicly by the LSST Science Pipelines, which also create a model of the profile. A mask for the satellite trail is created by forming the union between the above threshold regions of the profile model and the detected pixels. 
    }
    \label{fig:diff_image}
\end{figure}

In Figure \ref{fig:diff_image} we show the difference image between the single visit warp and a template created by assembling several aggressively clipped single visit exposures. The satellite trail is detectable and clearly visible by eye. The trail's corresponding pixels in the mask plane will be set to CLIPPED, and these pixels will not be propagated forward when producing a coadd.

\begin{figure}
    \centering
    \includegraphics[width=\columnwidth]{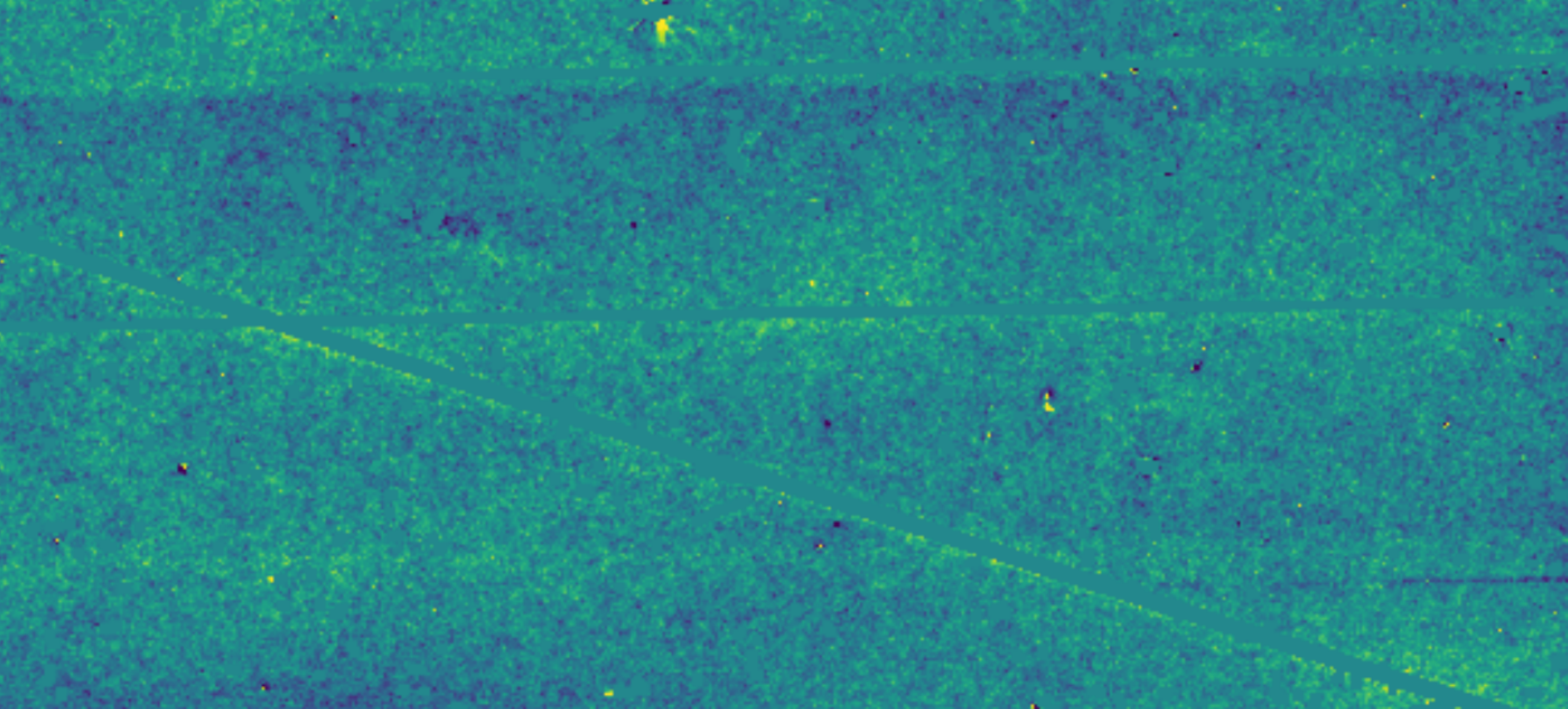}
    \caption{The same difference image from Figure \ref{fig:diff_image}, now with the satellite trail masked. Some low surface brightness features of the line profile bleed beyond the mask's edge in the difference image. 
    }
    \label{fig:diff_image_mask}
\end{figure}

In Figure \ref{fig:diff_image_mask} we show the same difference image, now with masking applied. Though the bulk of the flux of the satellite trail is masked out, some flux from the low surface brightness features of the extended tails of the satellite trail are still visible just on the edges of the mask.

\subsection{Examining trails in Coadds}
\begin{figure}
    \centering
    \includegraphics[width=\columnwidth]{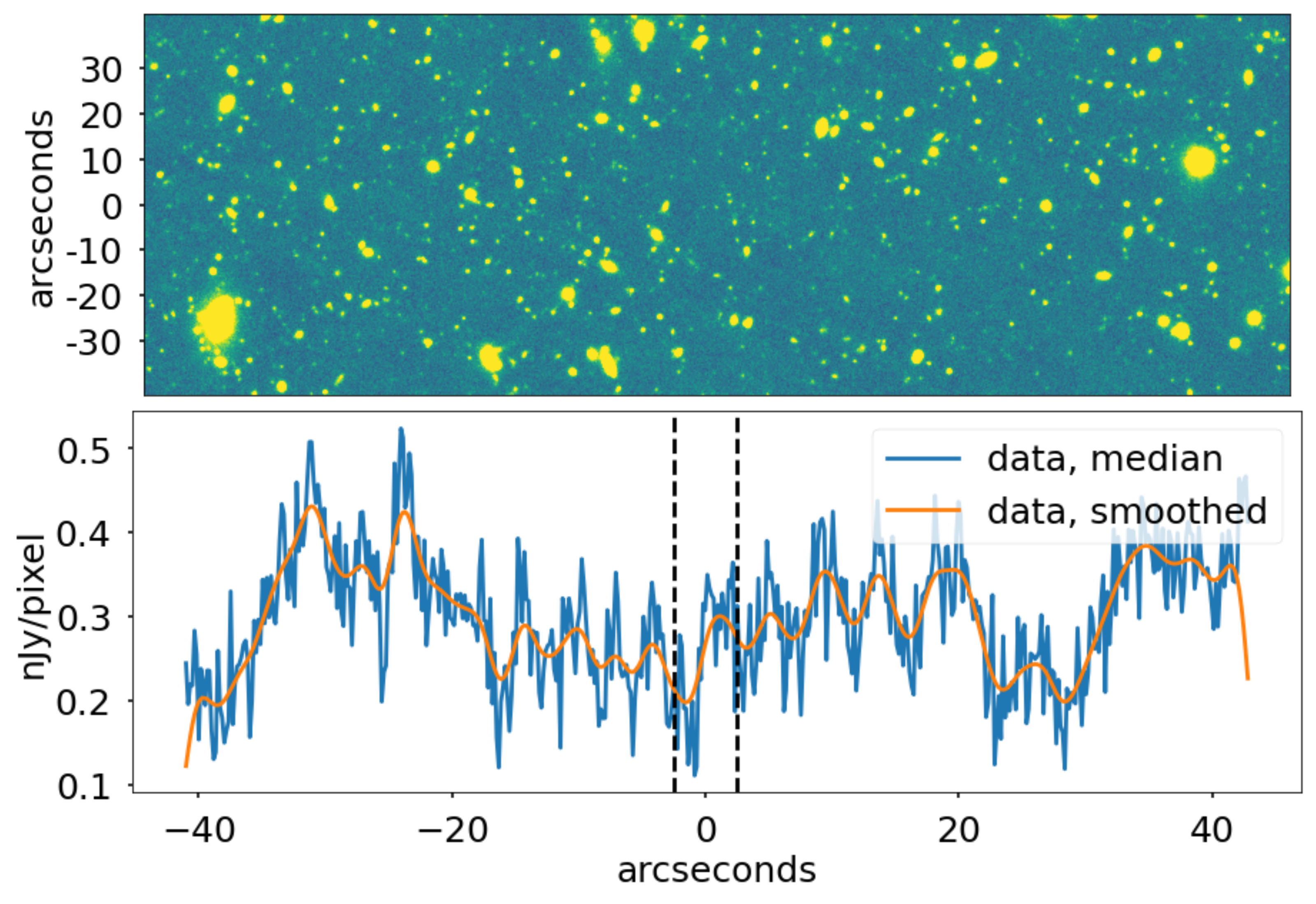}
    \caption{\emph{Top}: This coadd frame directly corresponds to the same location on the sky as Figure \ref{fig:warp}. 33 re-sampled warps were used to assemble this coadd frame, the warp shown in Figure \ref{fig:warp} included amongst them. The satellite trail is not easily visible by eye in the coadd. \emph{Bottom}: Surface brightness profile of the coadd at the location of the satellite trail in Figure \ref{fig:warp} collapsed along the x-axis, along with the approximate width of the mask displayed in dashed black lines. In orange we show the result of convolving the original signal in blue with a Gaussian filter, whose width is twice that of the Point Spread Function (PSF). 
    }
    \label{fig:coadd}
\end{figure}

Ultimately we wish to see if the coadd is affected by the satellite trail which appeared in one single visit warp. To assess this, we zoom in on the region of the coadd which corresponds to the region of sky shown in Figure \ref{fig:warp} where the satellite trail appears. In the top panel of Figure \ref{fig:coadd} we show the coadded frame. The satellite trail is not readily visible by eye. In the bottom panel of the Figure, we show the surface brightness profile where the satellite trail appeared in the single visit warp, over a scale of 80 arcseconds collapsed along the x-axis in the top panel. We also show in dashed black lines the approximate extent of the CLIPPED mask that is set by the coaddition process. In orange, we show the result of smoothing the original signal by a Gaussian filter whose width is twice that of the PSF. We would not expect to see the effect of the incompletely masked trail in this plot because of the dominance of relatively bright objects.

\begin{figure}
    \centering
    \includegraphics[width=\columnwidth]{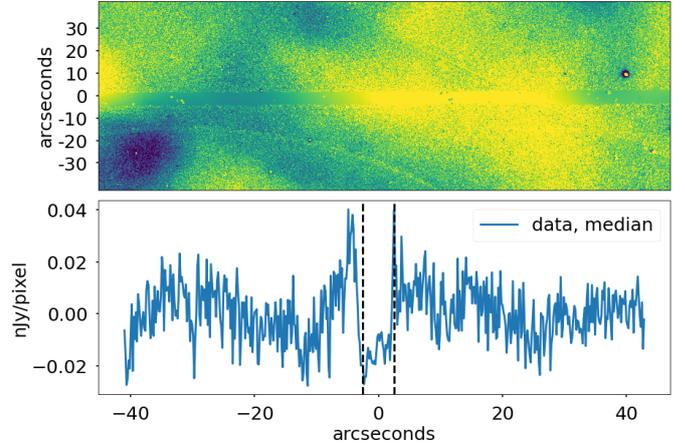}
    \caption{\emph{Top}: The difference between a coadd assembled using all individual visits-including the individual exposure with a satellite trail-and a coadd assembled with all visits \emph{except} the visit with a satellite trail. The default grown mask and some optical ghosts are readily visible in the difference image. \emph{Bottom}: The surface brightness profile of the above image orthogonal to the trail, centered on the location of the trail. The un-clipped wings of the flux profile of the satellite trail are visible in the profile.}
    \label{fig:diffim}
\end{figure}

We also compare coadds with and without the single visit that contains a satellite trail. We take the original coadd-consisting of 33 single visit warps including the single visit warp with a satellite trail-and compute the difference with a coadd made up of 32 single visit warps, this time excluding the single visit warp that contains the satellite trail. The difference image is shown in Figure \ref{fig:diffim} in the upper panel. The residual image shows the satellite trail default grown mask along with some optical ghosts. The origin of the smooth arcminute scale low surface brightness residuals may be that each visit is background subtracted using the skyCorr model prior to warping and coaddition. The skyCorr sky frame is quite different as a function of location on the focal plane. As a result, we expect small differences when including/excluding certain visits.

Fitting these out with a 4th degree 1-d polynomial, in the lower panel, we show the median surface brightness profile of the image orthogonal to the trail centered on the satellite trail's mask. There is a noticeable dip from the mask.  The spillover light from the satellite trail wings has an amplitude of about 0.03 nJy, which is barely significant in the sense that the resulting linear string of correlated pixels may generate a false shear signal in a catalog containing the faintest detected galaxies at the survey surface brightness limit (Kelvin et al 2021, in prep.).

There are failure modes in the LSST Science Pipelines (v2.1) where morphological filtering is not engaged where satellite trails are not masked correctly. These include satellite trails that pass through bright stars. Such unmasked satellite trails can propagate to the coadd. We discuss trail intensity fitting below. First, however, we examine the effects of varying mask widths.

\begin{figure}
    \centering
    \includegraphics[width=\columnwidth]{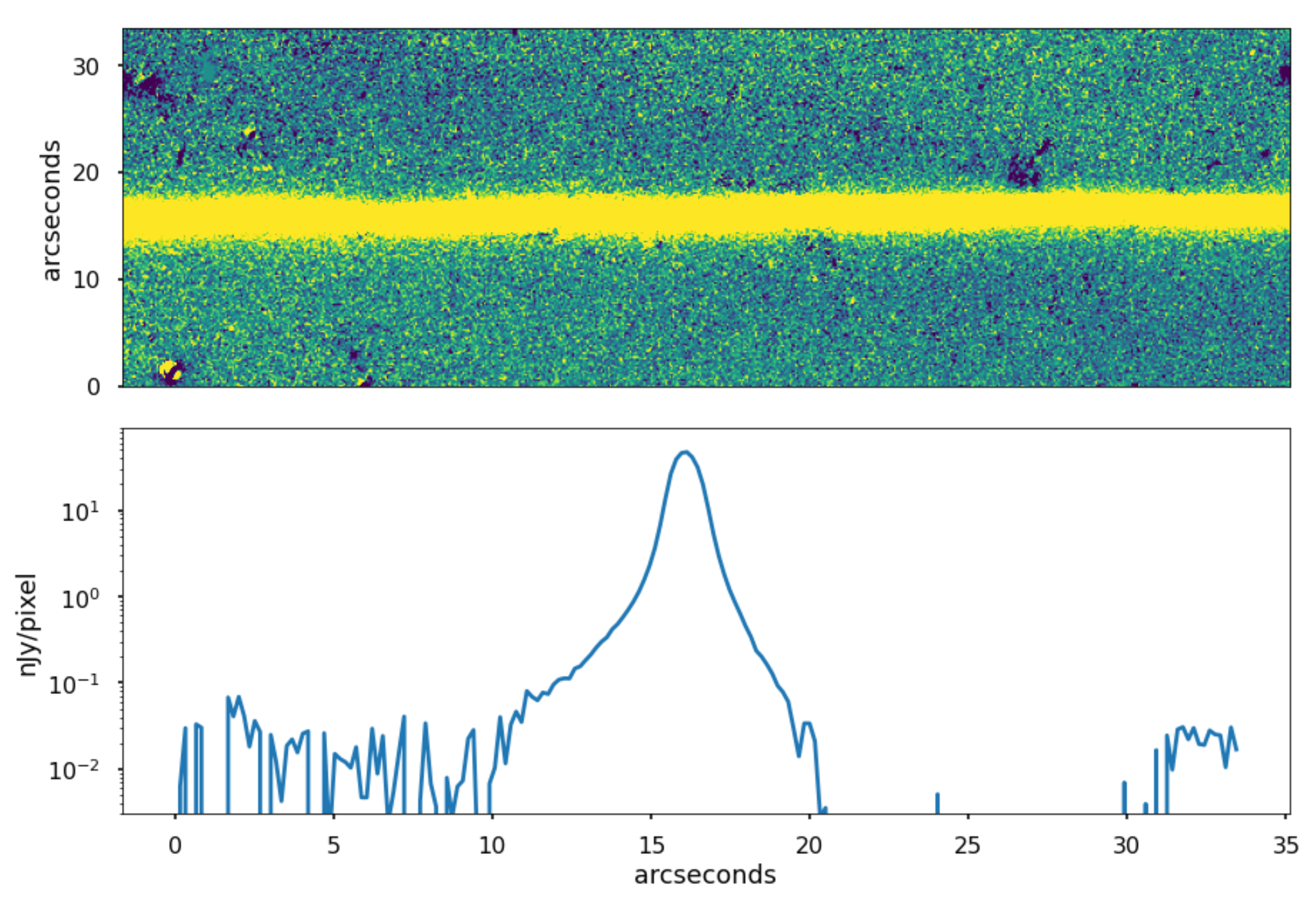}
    \caption{\emph{Top}: A zoom in of a difference image between a coadd which consists of 29 single visits (including a single visit which contains a satellite trail) and a coadd assembled using 28 single visits, \emph{excluding} the single visit image that contains a satellite trail. The standard coaddition process does not remove the satellite trail in the 29 image coadd. The differencing largely removes sources other than the satellite trail. \emph{Bottom}: The surface brightness profile of the above image, averaged in the direction orthogonal to the satellite trail. The sky level on the right hand side of the profile is suppressed, likely due to over subtraction of the background sky.}
    \label{fig:my_label}
\end{figure}

\section{Effectiveness of various mask widths}

\begin{figure}
    \centering
    \includegraphics[width=\columnwidth]{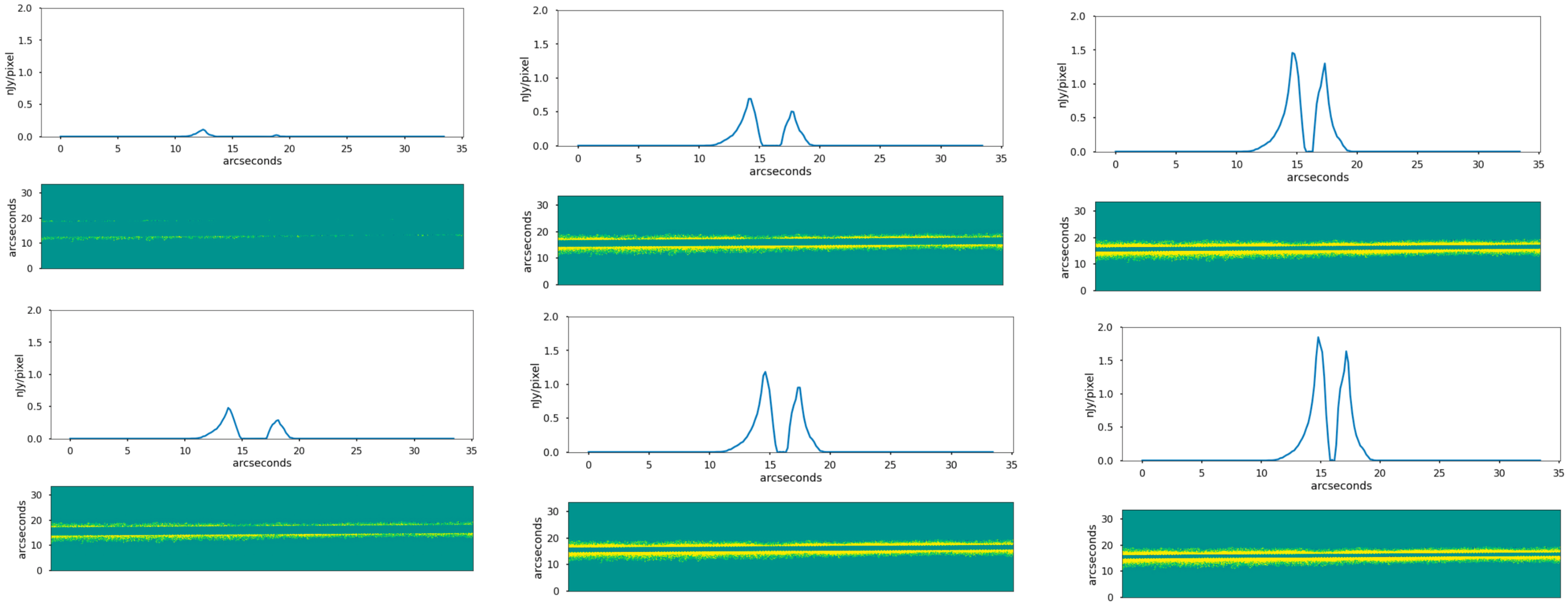}
    \caption{The spillover light in the coadd beyond masks of 6 different widths, shown on a linear flux scale which emphasizes the inner brightest region at the mask edge. }
    \label{fig:6masks}
\end{figure}

It is important to mask the satellite trails' surface brightness out to a level where the spillover light is below a surface brightness level which would cause systematic errors in the analysis of the coadd data. One can set parameters by hand or leave them in their default settings for the morphological filtering routine. With the default settings of the routine enabled, the routine is successful in identifying satellite trails that would otherwise go unmasked during the coaddition process. The masking of the trail occurs in the assembly of the coadd. In order to gauge the effectiveness of masking as a function of mask size, we study the satellite trail spillover light in the coadd beyond masks of various mask width. To do this, we first identify a coadd in the HSC COSMOS field where a satellite trail was missed by the initial pass of masking, but is successfully handled by the morphological filtering routine. The coadd in patch 4,8 for tract 9813 is such a coadd, consisting of 29 individual exposures. One of the 29 individual exposures contains a satellite trail, which passes over a chip gap. Without morphological filtering, this satellite trail will propagate to the coadd. However, enabling morphological filtering with its default settings is successful in masking the trail such that the surface brightness profile at its location in coadd cannot easily be distinguished from sky background noise. For the purposes of our study, we create several other coadds where we enable morphological filtering and adjust the parameter \texttt{maskStreaks.footprintThreshold}. This has the effect of adjusting the mask width to different widths in each coadd. The  setting does not necessarily map linearly to the mask width. Bearing this in mind, we iterate over values of \texttt{maskStreaks.footprintThreshold} to achieve an array of mask widths, knowing they may not sample mask widths in even intervals. We intentionally adjust the \texttt{maskStreaks.footprintThreshold} such that the trail is insufficiently masked, so we may study the effect of spillover light. Six examples of the remaining light after masks of 6 widths are shown in Figure \ref{fig:6masks} on a linear flux scale. The low surface brightness tail of the spillover light cannot be seen on a linear flux scale.

\begin{figure}
    \centering
    \includegraphics[width=\columnwidth]{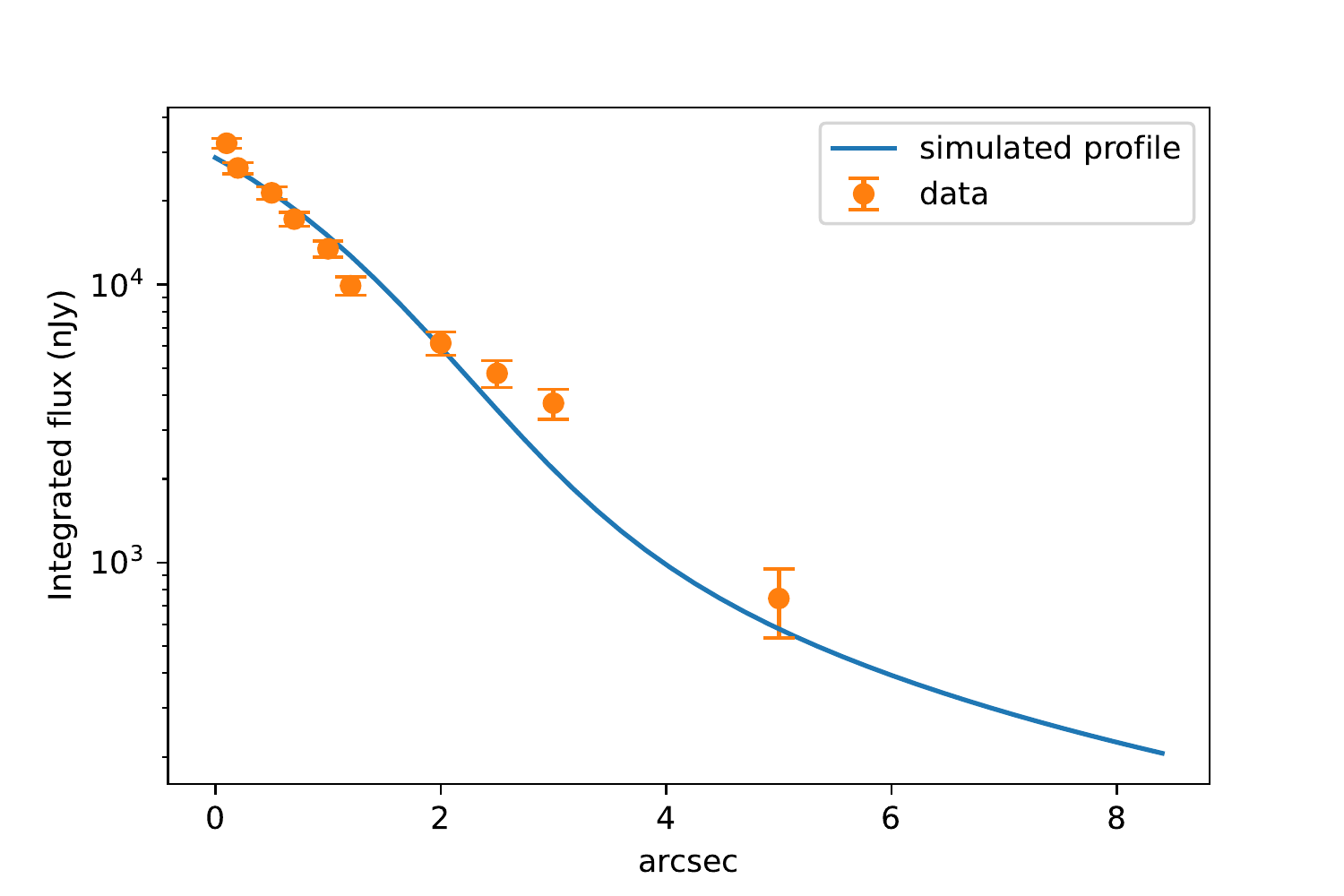}
    \caption{The integrated spillover light beyond masks as a function of mask width. Orange points are the data for ten mask widths, and the blue curve is a full atmosphere plus optics model fit to the data.}
    \label{fig:fluxfit}
\end{figure}

Once the coadd is created with the trail partially masked, we integrate the flux in excess of background over a large area along the trail and out away from the trail to the edge of the image. This is shown in Figure \ref{fig:fluxfit}, where the orange points represent the spillover integrated flux in the coadd beyond a mask of the indicated size in arcseconds.  For example, the point at 5 arcsecond mask width is the flux density (surface brightness) integrated out beyond the trail to the edge of the image -- total spillover light in the coadd. The code used to create this Figure can be found at 
\url{https://github.com/ih64/coadd4-8/blob/main/Patch4,8.ipynb }.

A model including the profile for the Starlink satellite convolved with a von Karman atmosphere \citep{ziad2000} for the observed airmass and the Subaru telescope optics, normalized to the peak near zero mask width, is shown in blue. Note that there is considerable flux in the coadd beyond a mask of 5 arcsecond width.

\begin{figure}
    \centering
    \includegraphics[width=\columnwidth]{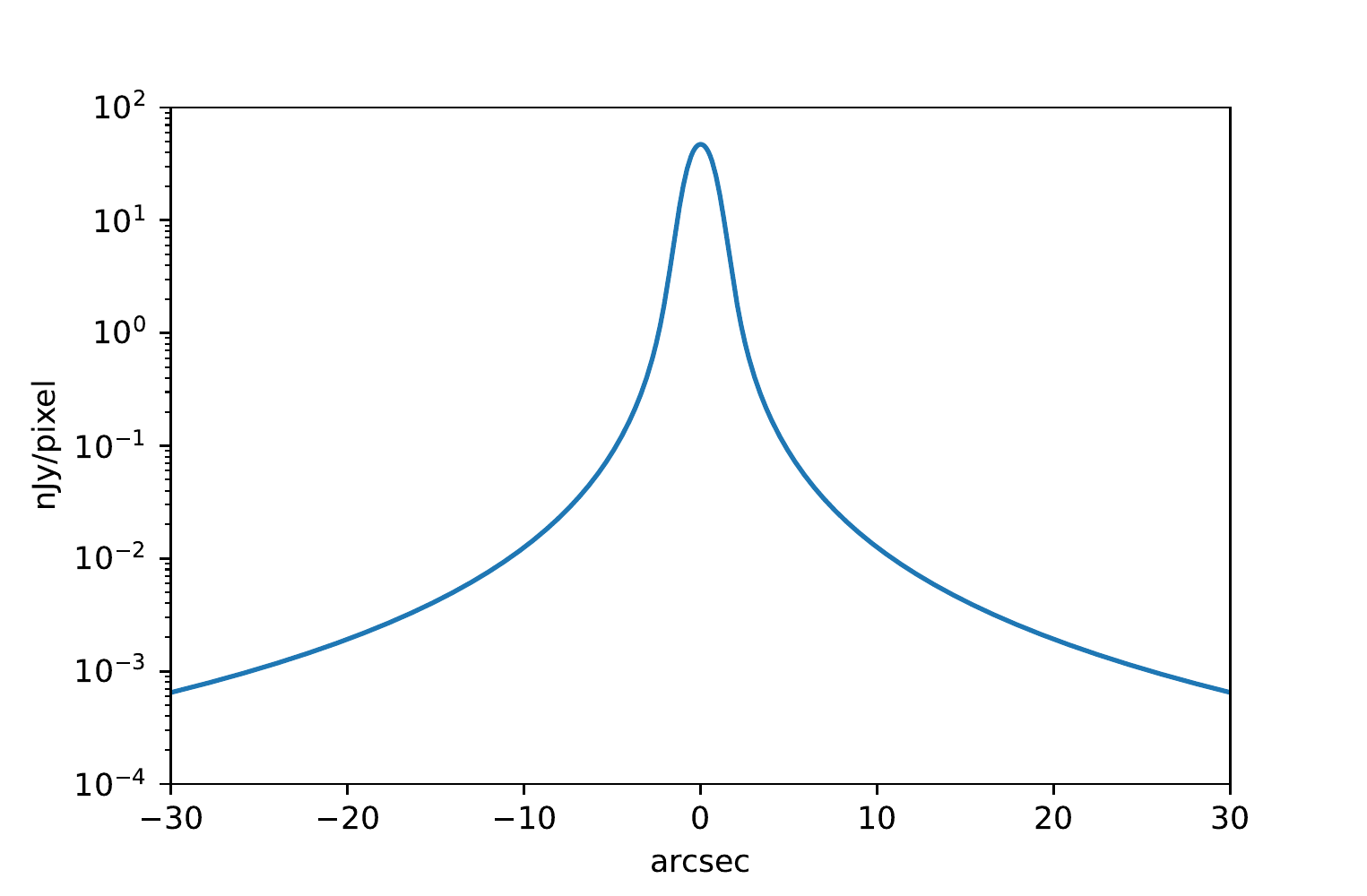}
    \caption{The surface brightness profile for the model fit to the mask data. This falls below the conservative threshold of 0.02 nJy/pixel for a mask width of 20 arcsecond.}
    \label{fig:sbfit}
\end{figure}

How wide a mask is needed in order to suppress the spillover light below the critical surface brightness limit for detectable systematics in the coadd?  A typical faint galaxy in the LSST 'Gold Sample' is $\sim$2 nJy/pixel, with surface brightness falling with radius. Since in galaxy shear estimates we must go out to arcsecond scale radii for 2nd moment shape measurement, the critical surface brightness is about 0.2 nJy/pixel.  Thus, we must examine the surface brightness of the spillover light out to distances from the satellite trail where it falls conservatively below a tenth of this level, or about 0.02 nJy/pixel.  In Figure \ref{fig:sbfit} we show the surface brightness profile of the model with shape parameters which were used in the integrated flux vs mask width study.

\begin{figure}
    \centering
    \includegraphics[width=\columnwidth]{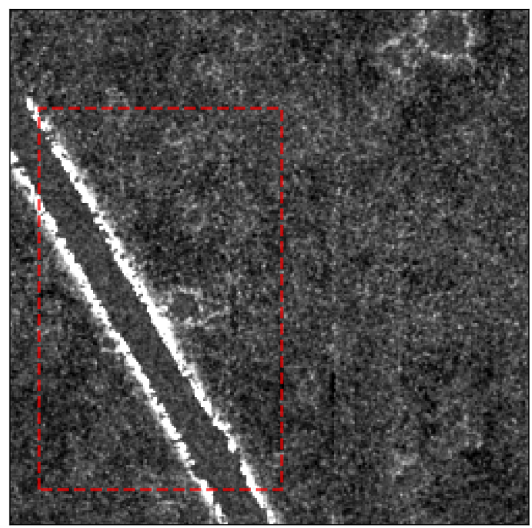}
    \caption{A LEO satellite seen in a r-band coadd image stack from DES DR2 observed using DECam on the Blanco telescope. A satellite trail in one of the 90 sec exposures was masked with 40-arcsecond-wide mask, and the coadd image was smoothed by a 1.3 arcsecond Gaussian kernel and rebinned by a factor of 10 to reveal low-surface-brightness features. Stars and galaxies were also masked, as can be seen. }
    \label{fig:decam}
\end{figure}

The broad wings of the atmospheric PSF cause the spillover light to extend to large distances from the satellite trail.  The theoretical fit for the surface brightness of this LEOsat falls below the conservative level of 0.02 nJy/pixel for the particular mask width 20 arcsecond. Brighter satellites will require correspondingly wider masks, and this is accounted for in the LSST Science Pipelines as the mask width is set by the brightness of the satellite trail. For example, a LEOsat two mag brighter would need a 40 arcsecond mask. There are satellite trail data from wide-field observatories other than Subaru.  Scaling from the 4m Banco to the effective 6.5m Rubin observatory, a 30 sec exposure in the LSST corresponds to a 90 sec exposure on the Blanco.  Recent Blanco DECam imaging has encountered LEO satellite trails. In a DES DR2 coadd of of many DECam exposures a satellite trail in one of them which was masked by a 40 arcsecond mask leaked though to the coadd as a bright spillover pair of trails\footnote{Alex Drlica-Wagner, private communication} straddling the masked trail. This is shown in Figure \ref{fig:decam}, where fainter spillover light from masked stars and galaxies may also be seen in the coadd. This image was first convolved with a Gaussian kernel with sigma = 5 pixels $\sim 1.3$ arcsecond and then binned by a factor of 10$\times$10 pixels ($\sim 2.6 \times 2.6$ arcsecond) in order to reveal low surface brightness. Clearly a wider mask is required at these surface brightness levels.  

\section{Beyond simple masking: model fit}

In order to suppress residual systematics printing through to the coadd, the masks have to be grown in width to reach beyond where the flux from the tails of the satellite trail reach a fraction of the sky surface brightness. Masks are binary (0,1), while the surface brightness variations are smooth on the scale of many pixels. To improve on masking with grown masks, a simple model fit to the trail surface brightness profile could be first subtracted.  Then the residuals along the peak of the trail could be masked with a much narrower mask.

\begin{figure}
    \centering
    \includegraphics[width=\columnwidth]{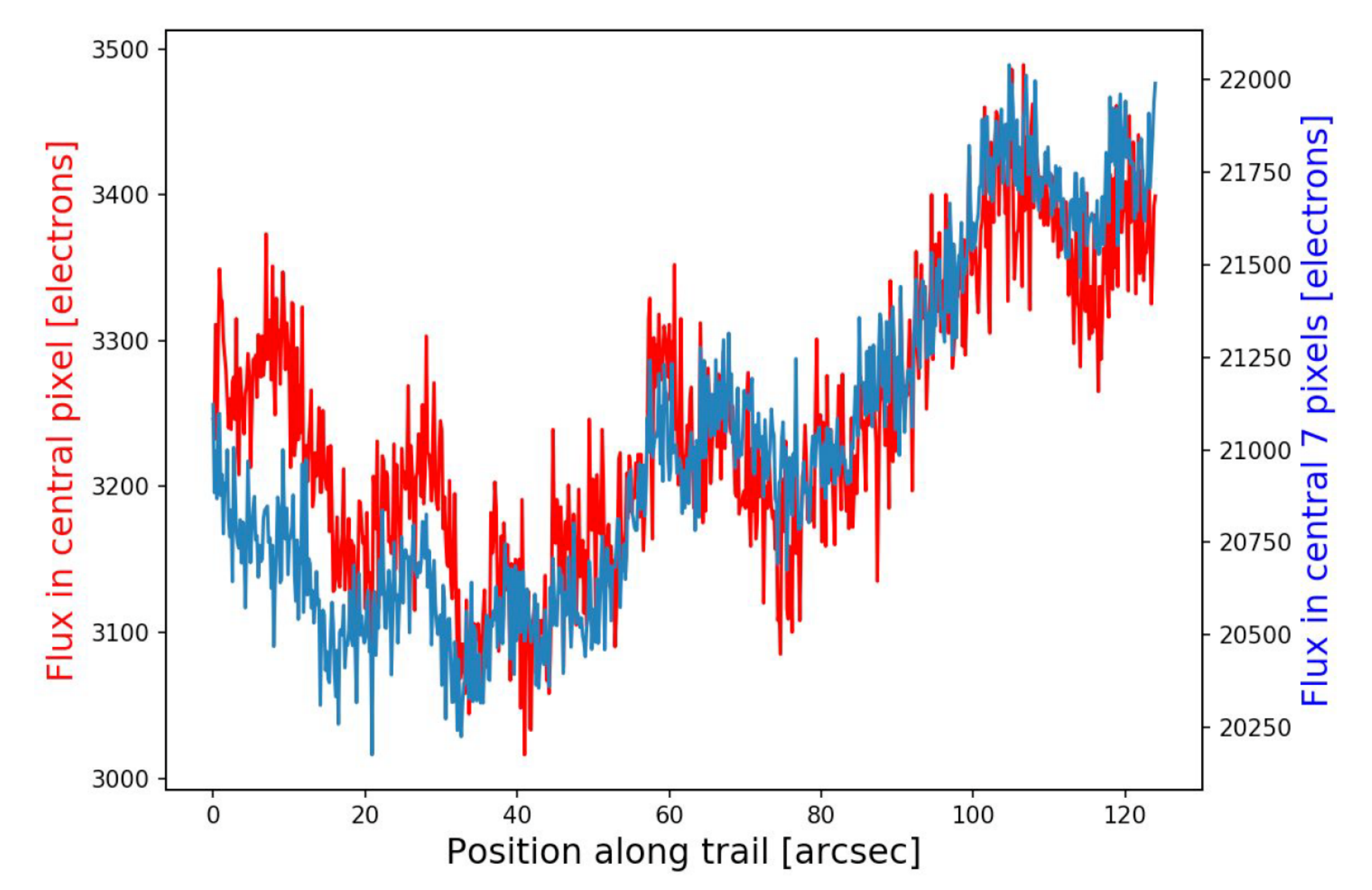}
    \caption{A satellite with MLI thermal blanket shows variations in flux along the trail. The blue curve is flux smoothed on the spatial scale of galaxies.  The result would be a linear string of bogus faint galaxies -- giving a systematic weak lens shear offset.}
    \label{fig:alongtrail}
\end{figure}

Of course this would be a longer term study, and there are potential issues. Due to variations in reflected sunlight due to changes in the solar angle during orbital passage, satellites exhibit fluctuations in flux along the trail on a variety of timescales, so the model fit would then have at least three fit parameters. The reflected solar flux in the direction of the observatory is not generally a continuous smooth function of time, but can vary rapidly due to the complex arrangement of surfaces of the satellite hardware. Satellites with MLI thermal blankets (such as OneWeb) exhibit flickering of their flux. An example is shown in Figure \ref{fig:alongtrail}, where variations in flux are seen on arcsecond spatial scales of galaxies. These considerations would set bounds on the model smoothing parameter along the trail direction, and would have to be optimized. 
Due to the sky noise: spillover flux causes linear features of correlated noise. Depending on the threshold for detection, this can produce a linear string of bogus faint galaxies near the noise level which may produce a systematic weak lens shear offset.  

\section{Possible impact on cosmology probes}

The impact on the cosmological probes of LSST will depend on several factors including the number of these linear bogus features leaking into the coadd, their intensity, and the number of satellites in the data. A forward simulation of these effects using a large N-body cosmology simulation is feasible, and we plan such a study. Due to the sharp cutoff at the mask edges, the majority of the spillover light is in a pair of narrow parallel noisy lines. This residual flux is spatially narrow because of the rapid decrease in spillover light orthogonal to the satellite trail. At the low surface brightness planned for LSST galaxies and cosmology science, for each satellite trail this gives rise to two parallel lines of faint bogus galaxies elongated along the trail direction.
Because each satellite trail will be in one band at a time, the print-through to the multiband static sky catalog will only occur at groups of pixels in excess of sky noise jointly at the same coordinates in all bands used for photometric redshift. As a result the systematic will appear only in the faint source sample galaxies, not in the brighter foreground lens sample. Thus any impact on cosmological probes would occur mainly in shear-shear correlations: cosmic shear.

\section{Summary}
We examine HSC data of the Hubble COSMOS field which are known to contain satellite trails in the individual exposures. We study the process by which these trails are masked in the LSST Science Pipelines v2.1. We find that the default grown masks generated by the LSST Science Pipelines remove most of the flux associated with the satellite trails in individual exposures. Beyond the masks, right at their edge, we find very low surface brightness tails of the satellite trail remain unmasked. These features, appear prominently in individual exposures, and appear at marginally significant levels in 33 image HSC coadds using the default grown mask. Additionally, we examine the effectiveness of the morphological filtering routine. We find that the routine in its default settings is successful in dealing with satellite trails that are otherwise missed on the initial pass of outlier rejection, masking most of the light. We also adjust the \texttt{footprintThreshold} parameter for the morphological filtering routine so we may study the effect of spillover light as a function of mask width. A model including the profile for the Starlink satellite convolved with a von Karman atmosphere is fit to this data. We observe in the fitted model that its broad wings extend to 20 arcseconds to surface brightness levels of 0.02 nJy/pixel. While this level of surface brightness is below the systematic threshold, brighter satellites will require broader masks.

Due to natural variations in satellite trail brightness with solar angle, a single mask width for a given satellite trail is not a good match. This leads to the need for a conservative wide mask and scaling the mask width for the brightness of each trail.
Finally, we consider the possibility of a model fit to the trail surface brightness profile along the trail, which may allow a more robust removal of spillover light and a significant increase in recovered sky area near the tail at low surface brightness.

\section*{Acknowledgements}

We thank Alex Drlica-Wagner and Caleb Levy for providing the DECam DES 6-year coadd image, and Leanne Guy for helpful comments. We acknowledge useful discussions with Adam Snyder, Andrew Bradshaw, Zeljko Ivezic, Craig Lage, and Erfan Nourbakhsh. This work was supported in part by NSF/AURA/LSST grant N5698CC.
The Hyper Suprime-Cam (HSC) collaboration includes the astronomical communities of Japan and Taiwan, and Princeton University. The HSC instrumentation and software were developed by the National Astronomical Observatory of Japan (NAOJ), the Kavli Institute for the Physics and Mathematics of the Universe (Kavli IPMU), the University of Tokyo, the High Energy Accelerator Research Organization (KEK), the Academia Sinica Institute for Astronomy and Astrophysics in Taiwan (ASIAA), and Princeton University. Funding was contributed by the FIRST program from the Japanese Cabinet Office, the Ministry of Education, Culture, Sports, Science and Technology (MEXT), the Japan Society for the Promotion of Science (JSPS), Japan Science and Technology Agency (JST), the Toray Science Foundation, NAOJ, Kavli IPMU, KEK, ASIAA, and Princeton University. 
This paper makes use of software developed for Vera C. Rubin Observatory. We thank the Rubin Observatory for making their code available as free software at http://pipelines.lsst.io/.

\bibliographystyle{elsarticle-num-names} 
\bibliography{Masking-study}
 








\end{document}